\documentclass[useAMS]{mn2e}
\usepackage{graphicx}
\usepackage{epsfig}
\usepackage{amssymb}
\usepackage{lscape}
\usepackage{ulem}
\usepackage{txfonts}

\def\aq{ALMaQUEST}

\def\sigh2{$\Sigma_{\rm H_2}$}

\def\sigsfr{$\Sigma_{\rm SFR}$}

\def\sigstar{$\Sigma_{\star}$}

\def\c2s{C\,{\sc ii}$^{\star}$}

\def\fgas{$f_{\rm H_2}$}

\title[ALMaQUEST VI - `retired' galaxy regions] {The ALMaQUEST Survey: VI.  The molecular gas main sequence of `retired' regions in galaxies.}

\author[Ellison et al.] {Sara L. Ellison$^1$,   Lihwai Lin$^2$, Mallory D. Thorp$^1$, Hsi-An Pan$^3$, Sebastian F. S\'{a}nchez$^4$, \newauthor  Asa F. L. Bluck$^6$, Francesco Belfiore$^7$\\ 
$^1$ Department of Physics \& Astronomy, University of Victoria, Finnerty Road, Victoria, British Columbia, 
  V8P 1A1, Canada\\
  $^2$ Institute of Astronomy \& Astrophysics, Academia Sinica, Taipei 10617, Taiwan\\
  $^3$ Max-Planck-Institut f\"ur Astronomie, K\"onigstuhl 17, D-69117 Heidelberg, Germany\\
  $^4$ Instituto de Astronom\'{i}a, Universidad Nacional Autonoma de Mexico, A. P. 70-264, C.P. 04510, Mexico, 
  D.F., Mexico\\
  $^6$ Kavli Institute for Cosmology \& Cavendish Astrophysics, University of Cambridge, Madingley Road, Cambridge, CB3 0HA, UK\\
  $^7$ INAF –- Osservatorio Astrofisico di Arcetri, Largo E. Fermi 5, I-50125, Florence, Italy
}

\begin{document}

\maketitle

\begin{abstract}
In order to investigate the role of gas in the demise of star formation on kpc-scales, we compare the resolved molecular gas main sequence (rMGMS: \sigh2 vs \sigstar) of star-forming regions to the sequence of `retired' regions that have ceased to form new stars.  Using data from the \aq\ survey, we find that retired spaxels form a rMGMS that is distinct from that of star-forming spaxels, offset to lower \sigh2\ at fixed \sigstar\ by a factor of $\sim$5.  We study the rMGMS of star-forming and retired spaxels on a galaxy-by-galaxy basis for eight individual \aq\ galaxies.  Six of these galaxies have their retired spaxels concentrated within the central few kpc. Molecular gas is detected in 40-100\% of retired spaxels in the eight galaxies in our sample.  Both the star-forming and retired rMGMS show a diversity in normalization from galaxy-to-galaxy.  However, in any given galaxy, the rMGMS for retired regions is found to be distinct from the star-forming sequence and gas fractions of retired spaxels are up to an order of magnitude lower than the star-forming spaxels.  We conclude that quenching is associated with a depletion (but not absence) of molecular gas via a mechanism that typically begins in the centre of the galaxy.
\end{abstract}

\begin{keywords}
Galaxies: ISM, galaxies: star formation, galaxies: evolution, galaxies: general
\end{keywords}

\section{Introduction}

The processes proposed to quench and regulate star formation in galaxies are manyfold.  The simplest of these potential pathways is via the mundane process of gas exhaustion, whereby past generations of stars have consumed all of the available fuel.  Alternatively, the gas could be forcibly removed, either via external, environmental processes such as ram pressure stripping, or driven by internal energy sources such as starbursts or an active galactic nucleus (AGN).  However, the relatively common presence of a persistent gas reservoir in galaxies normally considered to be quiescent, indicates that the absence of gas may not be the dominant route to quenching (e.g. Young et al. 2011; Thom et al. 2012;  French et al. 2015; Davis et al. 2019).    Moreover, the processes described above appear to have a complex relationship with the regulation of star formation; whilst they have the potential to quench star formation, both stripping (e.g. Vulcani et al. 2018) and AGN (e.g. Carniani et al. 2016) have been observed to similarly foster new star formation.  It has also been found that most outflows do not possess sufficient energy to evacuate the gas from its host (e.g. Roberts-Borsani \& Saintonge 2018; Fluetsch et al. 2019).  Alternatively, galaxies may retain their gas, but find themselves unable to convert it into stars, e.g. due to high stability or turbulence in the interstellar medium (e.g. Martig et al. 2009; Mendez-Abreu et al. 2019; Gensior et al. 2020).  The relative importance of these processes may also depend on redshift; in the current paper we focus on quenching in nearby ($z<0.1$) galaxies.

Studying the spatial distibution of star formation within nearby galaxies has the potential to disentangle these various potential quenching mechanisms.  Using statistical samples of galaxies mapped with integral field unit (IFU) surveys, it has been shown that galaxies with low global star formation rates (SFRs) exhibit suppressed \sigsfr\ at all radii, with the largest deficits in the inner regions (Belfiore et al. 2018; Ellison et al. 2018; Medling et al. 2018; Wang et al. 2019).  Such inside-out quenching hints at processes linked to the inner galactic regions, a scenario supported by the observed correlation between central galactic properties (such as central mass density, central velocity dispersion and bulge mass) and quenched fraction (Lang et al. 2014; Omand et al. 2014; Bluck et al. 2014, 2016, 2020a).  However, environmental quenching seems also to play a role for galaxies located in densely populated neighbourhoods, where star formation seems to be quenched from the outside-in (Schaefer et al. 2017; Bluck et al. 2020b).

Further clues to the mechanisms that drive quenching may be gleaned by mapping the molecular gas content of galaxies in the process of ramping down their star formation.  A first step in this direction was made by Lin et al. (2017), who presented Atacama Large Millimeter Array (ALMA) maps of CO(1-0) obtained on the same spatial scale as optical IFU maps from the Mapping Nearby Galaxies at Apache Point Observatory (MaNGA) for three green valley galaxies.  Lin et al. (2017) (and follow-up work by Brownson et al. 2020) found low bulge gas fractions in their green valley sample, indicating that depleted inner gas reservoirs could be the cause of reduced SFRs.  Using indirect measurements of \sigh2\ (based on a calibration of spaxel-by-spaxel dust attenuation) Sanchez et al. (2018) also found a deficit of molecular gas in the central regions of green valley galaxies.

In the work presented here, we take a complementary approach to investigating how the molecular gas content is linked to quenching on kpc-scales.  Rather than selecting galaxies via their global SFRs (e.g. those in the green valley), we take a broader view, and select any galaxy that exhibits detectable regions of internal quenching.  This more inclusive approach can identify galaxies that are broadly still forming stars, and hence has the potential to catch the cessation of star formation at an earlier stage.  The molecular gas properties of star-forming and quenching regions within a given galaxy are then assessed by comparing the resolved molecular gas main sequences (rMGMS, i.e. \sigstar\ vs. \sigh2) of the two internal populations on a galaxy-by-galaxy basis.



\section{The ensemble resolved molecular gas main sequence of retired spaxels}\label{ens_mgms_sec}

The data used in the work presented here are taken from the \aq\ survey\footnote{http://arc.phys.uvic.ca/$\sim$almaquest/} (Lin et al. 2020).  The \aq\ survey consists of 46 galaxies selected from the MaNGA survey that have been observed with ALMA in order to obtain CO(1-0) maps on the same spatial ($\sim$kpc) scale.  The galaxy sample, data acquisition, reduction and mapping to the MaNGA spatial grid are described in detail in Lin et al. (2020).  As in our previous papers (Ellison et al. 2020a,b), we use extinction corrected emission line fluxes and inclination corrected \sigstar\ from PIPE3D (Sanchez et al. 2016; 2018).  We refer the reader to those works for further details on all of these facets of the sample and data processing.  

The identification of galactic regions that are either star-forming or in the process of quenching is achieved via the use of line ratio diagrams (e.g. Baldwin, Phillips \& Terlevich 1981; hereafter BPT) and an H$\alpha$ equivalent width (EW) cut.  Regions which are no longer actively forming new stars, and whose emission lines are photoionized by aging stellar populations have low H$\alpha$ EWs, and have hence been dubbed as `retired' (Stasinska et al. 2008; Cid-Fernandes et al. 2010; Sanchez et al. 2014).  Following these previous works, spaxels are classified as star-forming or retired using the Kauffmann et al. (2003; hereafter K03) line ratio diagram according to the following criteria:

\begin{itemize}
\item  Star-forming:  Spaxels have S/N$>$2 in [NII], H$\alpha$, [OIII] and H$\beta$, lie below the K03 AGN demarcation and have H$\alpha$ EW $>$ 6 \AA.
\item  Retired:  Spaxels have S/N$>$2 in [NII], H$\alpha$, [OIII] and H$\beta$ and have H$\alpha$ EW $<$ 3 \AA.
\end{itemize}

\begin{figure}
	\includegraphics[width=8.5cm]{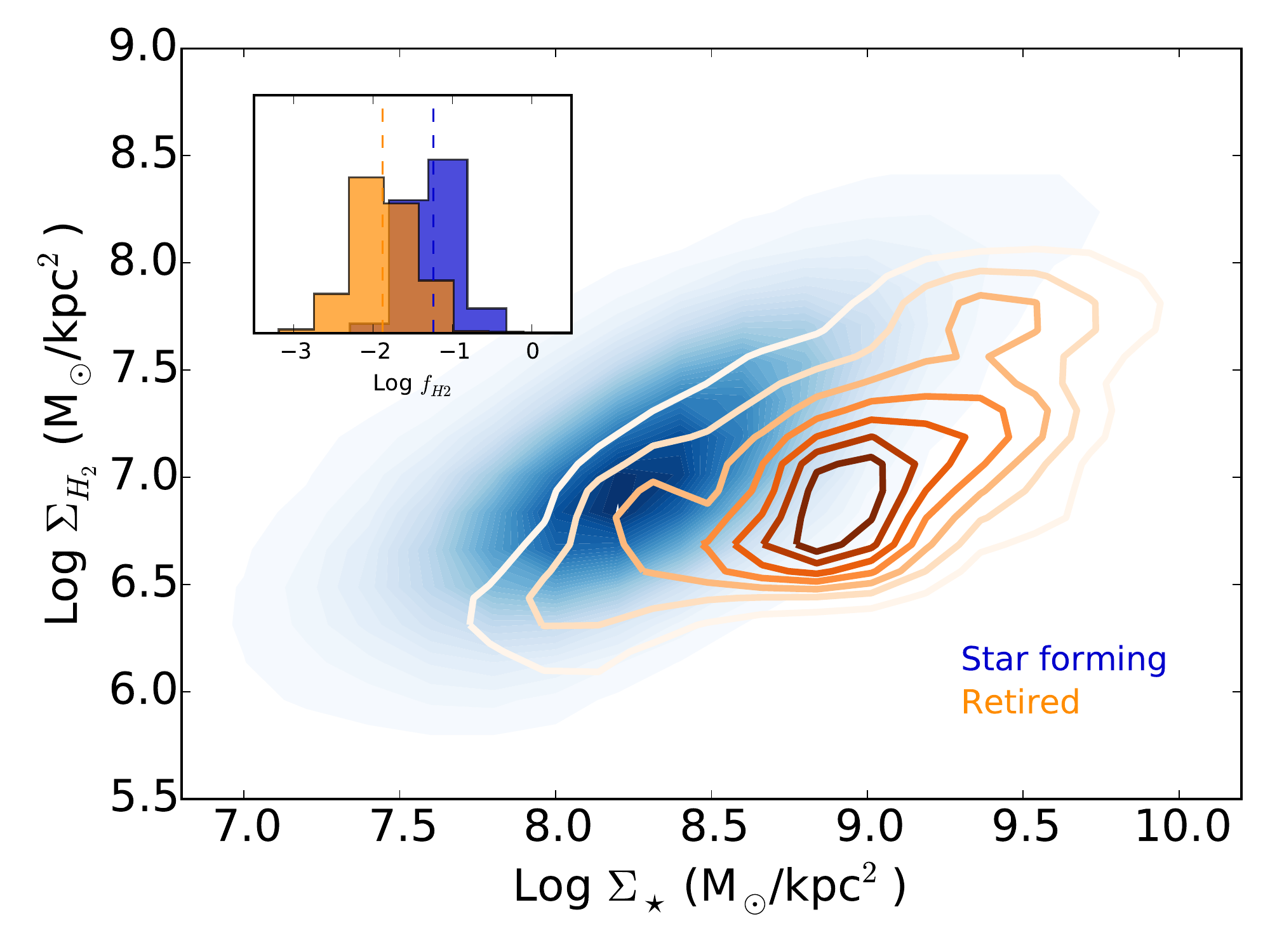}
        \caption{The resolved molecular gas main sequence for $\sim$19,000 star forming spaxels (blue shading) and $\sim$1,500 retired spaxels (orange contours) in the full \aq\ sample (46 galaxies).    Whereas the retired spaxels form their own version of a rMGMS, it is offset to lower \sigh2\ at fixed \sigstar.  The inset histogram shows the (normalized) distribution of molecular gas fractions of the two populations.  The (logarithmic) median gas fractions (shown by vertical dashed lines) of the star forming and retired populations are $-1.2$ and $-1.9$ dex, respectively. The detection threshold for \sigh2\ varies from galaxy-to-galaxy, but is typically log \sigh2 $\sim$ 6.3 -- 6.8 M$_{\odot}$kpc$^{-2}$.}
    \label{MGMS_ensemble}
\end{figure}

The reason that we do not enforce a non-AGN classification on the BPT diagram for retired spaxels, which are simply characterized by weak H$\alpha$ emission, is that the majority of them lie above the AGN demarcation (e.g. Stasinska et al. 2008; Cid-Fernandes et al. 2011), despite not being photoionized by nuclear accretion processes.  We have further checked that the conclusions of this work are not sensitive to the requirement that retired spaxels be detected in other emission lines, or on the details of S/N threshold. Spaxels selected in the same way as our retired sample are sometimes referred to in the literature as LI(N)ERs (e.g. Belfiore et al. 2017; Hsieh et al. 2017).  In this work, we prefer the term `retired', to reflect that the photoionization of these spaxels is believed to originate from old stellar populations.  We note that the above spaxel classifications do not include all spaxels, e.g. those with intermediate H$\alpha$ EWs, or insufficient S/N in all four of the emission lines required for the K03 classification.  These spaxels do not receive a photoionization classification in this work.  

\begin{figure*}
	\includegraphics[width=16cm]{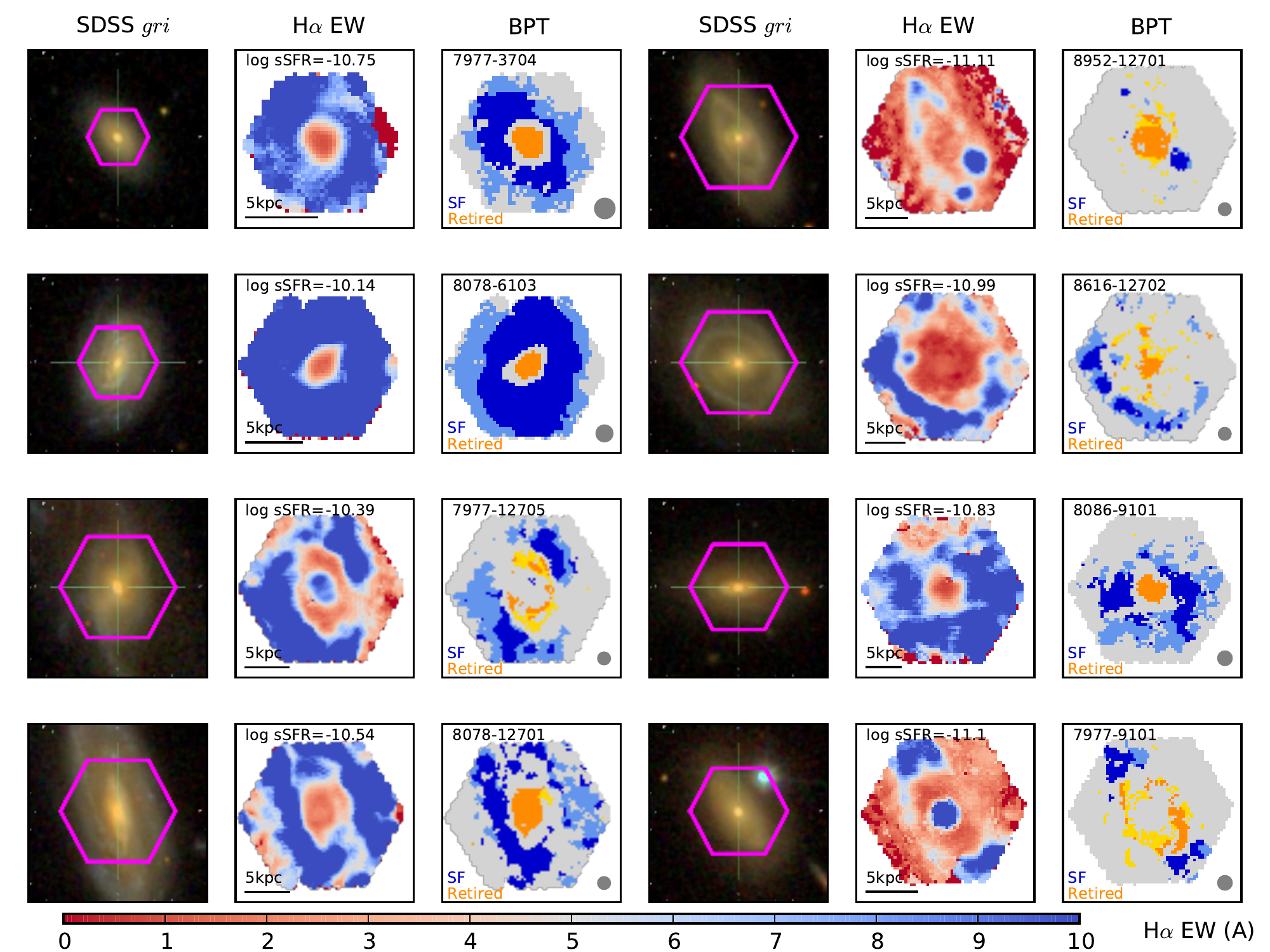}
        \caption{The 8 galaxies selected from the \aq\ sample which have sufficient ($>$20) spaxels to investigate the rMGMS for star-forming and retired populations of spaxels.  The first and fourth columns show the SDSS images with the hexagonal MaNGA footprint shown in magenta; the second and fifth columns show the H$\alpha$ EW maps (see colour bar along the lower edge of the figure); the third and sixth columns show the maps of spaxel classifications.  Retired spaxels are shown in orange and star-forming spaxels in blue, with paler shades of each colour indicating a non-detection in CO.  Spaxels that do not fall into either of these two classes are shown in grey.  The filled grey circle indicates the median FWHM of 2.5 arcsecond (Law et al. 2015).  
        }
    \label{map_fig}
\end{figure*}

In order to study the rMGMS, we additionally require a CO(1-0) detection which is converted to a \sigh2\ assuming a constant $\alpha_{CO}$=4.3 M$_{\odot}$ pc$^{-2}$ (K km s$^{-1}$)$^{-1}$.  From the spaxels in the star-forming and retired categories described above, we therefore further select spaxels with S/N$>2$ in their CO(1-0) line.  There are approximately 19,000 star-forming and 1,500 retired spaxels thus selected (although over-sampling means that these are not independent).  Although it would be of great interest to additionally study the location of quenched spaxels (characterized by lack of H$\alpha$ emission) relative to the star-forming rMGMS, there are only $\sim$100 quenched spaxels with CO detections in the entire \aq\ survey, prohibiting a statistical assessment.

In Fig. \ref{MGMS_ensemble} we show the rMGMS for the star-forming (blue shading) and retired spaxels (orange contours) for the ensemble of spaxels selected from the full \aq\ survey (46 galaxies). The star-forming spaxels show the familiar tight sequence shown by previous works (e.g. Lin et al. 2019; Morselli et al. 2020; Ellison et al. 2021), although we note that the absence of points in the lower right of the diagram is (at least in part) due to imposed detection thresholds.  The retired spaxels also exhibit a correlation between \sigstar\ and \sigh2, although offset to low \sigh2\ at fixed \sigstar\ compared with the sequence of star-forming spaxels (see also Sanchez et al. 2020, but using indirect \sigh2\ measurements).  The inset histogram in Fig. \ref{MGMS_ensemble} shows the molecular gas fraction (defined as log \fgas\ = log \sigh2\ - log \sigstar) distributions of the two populations. The median gas fraction of retired spaxels is a factor $\sim$5 lower than that of star-forming spaxels.   Therefore, Fig. \ref{MGMS_ensemble} hints that the low sSFRs in retired spaxels might be due to lower molecular gas fractions, on average.  An $\alpha_{CO}$ that increases with SFR (e.g. Accurso et al. 2017) would only increase the difference between the rMGMS of the two populations.

As we will show later in this Letter (see also Belfiore et al. 2017, 2018; Sanchez et al. 2018) many of the regions that host retired spaxels are in the centres of galaxies.  The population of retired spaxels is therefore naturally shifted to higher values of \sigstar, on average.   However, Fig. \ref{MGMS_ensemble} shows that the rMGMS of the retired spaxels is not simply an extension of the star-forming sequence at higher \sigstar.  The offset between the rMGMS of retired and star-forming spaxels is therefore not simply a radius effect.  Instead, Fig. \ref{MGMS_ensemble} suggests that the lower sSFRs of retired spaxels are linked to a deficit of molecular gas.

\section{Galaxy-by-galaxy resolved molecular gas main sequence of retired spaxels}\label{gbyg_mgms_sec}

Ellison et al. (2021) have shown that galaxy-to-galaxy variation dominates the scatter in the star-formation scaling relations.  To further dissect the rMGMS of retired galaxies, we therefore investigate the rMGMS on a galaxy-by-galaxy basis.

We begin by selecting only the 33 \aq\ galaxies with axial ratios $b/a>0.35$, based on inclinations provided in the NASA Sloan Atlas (NSA).  This sample is further reduced by requiring that at least 20 star-forming and 20 retired spaxels be present in a given galaxy, in order to determine the rMGMS for these two populations (in practice, the minimum number of retired spaxels in a given galaxy is 50).  In Fig. \ref{map_fig} we show the eight galaxies that remain after the axial ratio and spaxel classification cuts are made, ordered by total stellar mass.  For each galaxy we show the SDSS image (1st and 4th columns), the H$\alpha$ EW (2nd and 5th columns) and spaxel classification (BPT) map (3rd and 6th columns).  In the BPT maps spaxels classified as star-forming are shown in blue and retired spaxels are shown in orange, with paler shades of each colour indicating a CO non-detection.  The MaNGA plate-ifu identifier is given in the top of each BPT map and the global sSFR (from the PIPE3D value-added catalog) is given in the top of the H$\alpha$ EW map.  Fig. \ref{map_fig} shows that many galaxies in our sample host significant star formation and have sSFR values that are within the scatter of the main sequence, demonstrating the complementarity of our approach to studies of globally identified green valley (or retired galaxies) galaxies (e.g. Lin et al. 2017; Sanchez et al. 2018; Brownson et al. 2020).

Belfiore et al. (2017) have demonstrated that the photoionization morphologies of galaxies with retired populations fall into one of two broad categories:  Centrally concentrated or extended (termed cLIERs and eLIERs in that work).  These same broad morphologies are seen in our sample.  Fig. \ref{map_fig} shows that the retired spaxels in 6/8 of the galaxies in our sample are centrally concentrated, often (but not always) surrounded by a spatially extended region of star formation (i.e. cLIERs in the terminology of Belfiore et al. 2017).   The remaining two galaxies (7977-12705 and 7977-9101) have high H$\alpha$ EWs in their central regions, with the retired spaxels arranged in a ring around.  However, for both galaxies these central high H$\alpha$ EW spaxels are not classified as star-forming as they are above the K03 demarcation, and would be traditionally be classified as `composite' as they are below the Kewley et al. (2001) boundary.   One of these galaxies (7977-12705) shows signs of a past interaction, which may have triggered both central star formation and AGN activity (e.g. Ellison et al. 2013).  

\begin{figure}
	\includegraphics[width=9cm]{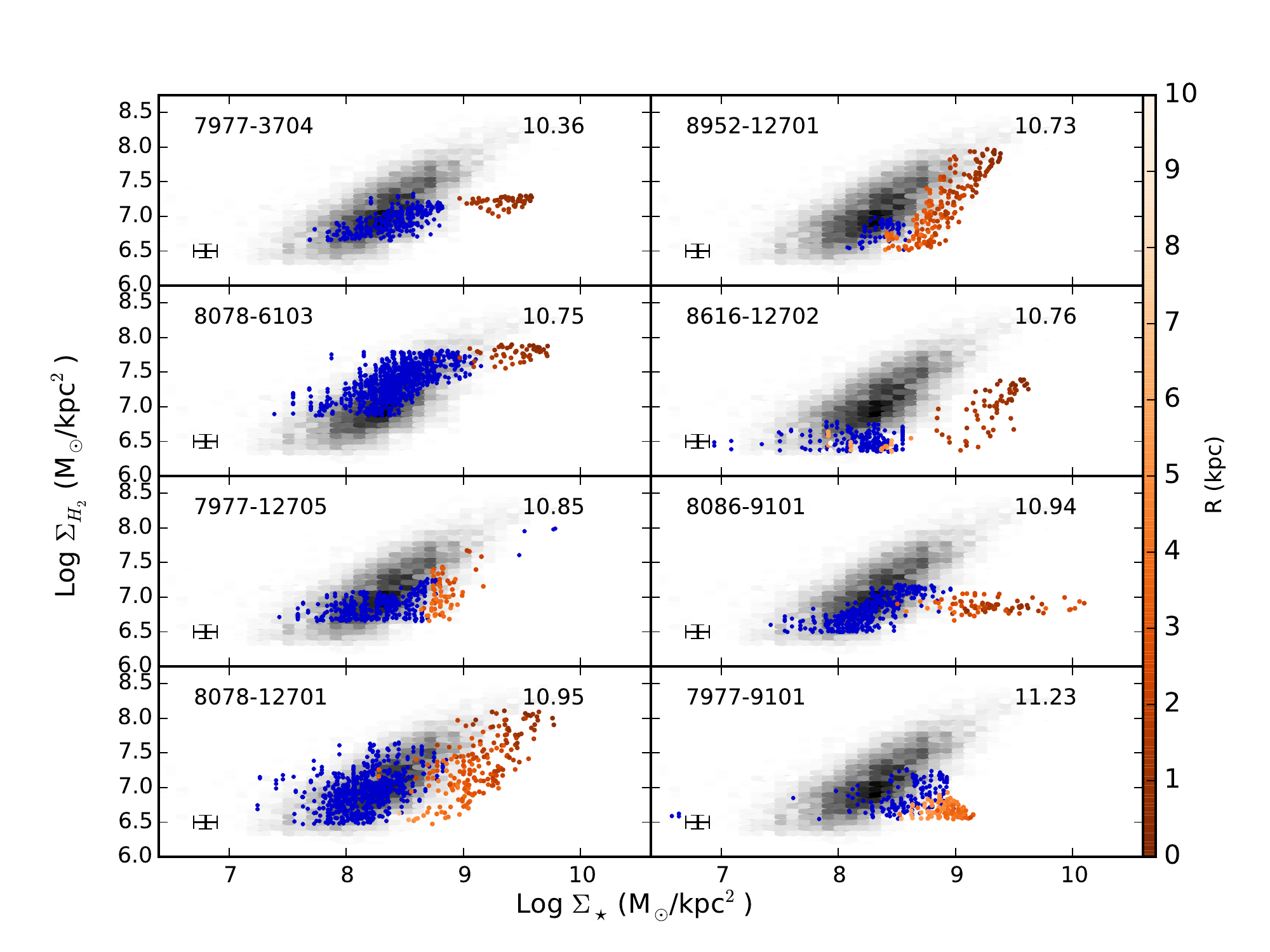}
        \caption{The rMGMS for eight \aq\ galaxies.  Star-forming spaxels are shown in blue and retired spaxels are shown in orange, with shading that indicates the distance from the centre of the galaxy in kpc.  The MaNGA plate-ifu is given in the top left of each panel and the stellar mass in the top right.  The retired spaxels show a distinct rMGMS compared with the star-forming spaxels in the same galaxy.  Typical measurement uncertanties are shown in the lower left of each panel; these do not include systematic uncertainties or modelling assumptions , e.g. due to choice of $\alpha_{CO}$, initial mass function, extinction law etc.}
    \label{mgms_pifu}
\end{figure}

In Fig. \ref{mgms_pifu} we show the rMGMS on a galaxy-by-galaxy basis for the eight galaxies selected above.  The panels appear in the same order as Fig. \ref{map_fig}; the MaNGA plate-ifu and total stellar mass are given in the top left/right of each panel.  The greyscale histogram is the same in each panel and contains all of the star-forming spaxels shown in Fig. \ref{MGMS_ensemble} for reference.  The rMGMS for the star-forming/retired spaxels in each galaxy is shown in blue/orange points, with shading further indicating their distance from the galactic centre.

Fig. \ref{mgms_pifu} demonstrates that, just as there is no universal (across all galaxies) rMGMS for star-forming spaxels (Ellison et al. 2021), so the rMGMS of retired spaxels exhibits large galaxy-to-galaxy variations.  The retired sequences are always offset from the star-forming sequences.  The retired sequences are characterized by high values of \sigstar; this is unsurprising given that in all eight galaxies (even the two whose retired spaxels form an extended ring) the retired spaxels are at smaller radii than the star-forming spaxels in the same galaxy.  However (perhaps with the exception of 7977-12705, which is a post-merger), the retired rMGMS is clearly not a simple extension to higher \sigstar\ of the star-forming rMGMS.   Fig. \ref{mgms_pifu} shows that the retired rMGMS in a given galaxy is gas-poor compared to the star-forming rMGMS extrapolated to high \sigstar.

\begin{figure}
	\includegraphics[width=9cm]{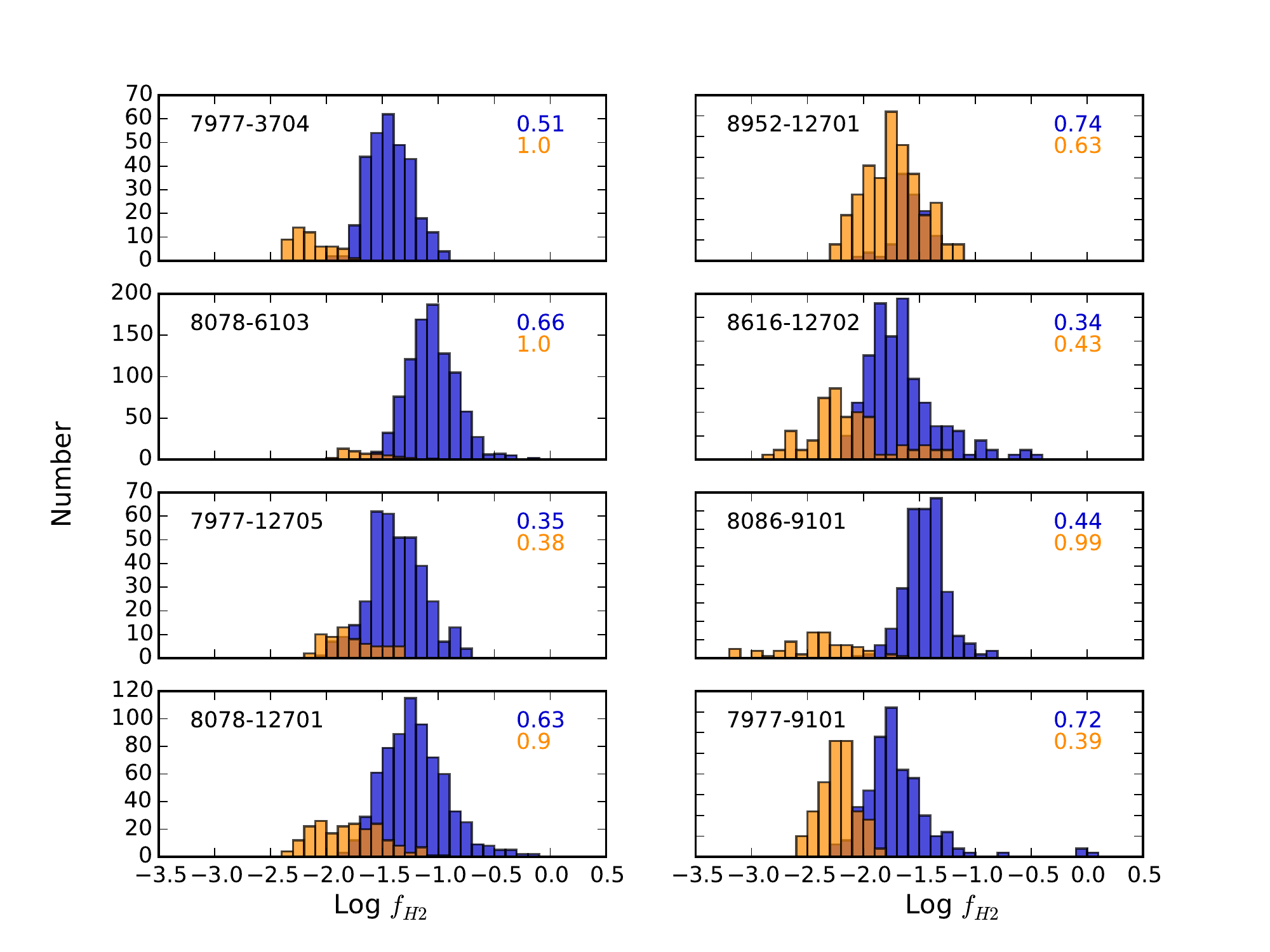}
        \caption{Gas fraction distributions for the eight galaxies in our sample.  The distribution for star-forming spaxels is shown in blue and for retired spaxels in orange.  The MaNGA plate-ifu is given in the top left of each panel and the fraction of star-forming/retired spaxels detected in CO (S/N$>$2) in the top right.  Panels appear in the same order as Fig. \ref{mgms_pifu}.  Retired spaxels have lower molecular gas fractions by up to an order of magnitude compared with star-forming spaxels in the same galaxy.}
    \label{fgas}
\end{figure}

We further examine the spaxel molecular gas fractions on a galaxy-by-galaxy basis in Fig. \ref{fgas}.  The fraction of star-forming/retired spaxels detected in CO is given in the top right corner of each panel and is 35-74 \% for star-forming spaxels and 38 - 100 \% for retired spaxels.  However, these dectection fractions are not directly comparable given the different radial (and hence \sigstar) ranges for the two populations.  A more informative comparison is therefore the gas fractions of the two populations.  The typical median gas fractions of (CO-detected) star-forming spaxels in a given galaxy is $\sim$ $-1.0$ to $-1.5$, whereas the retired spaxels are up to an order of magnitude more gas poor.  Therefore, whilst it is possible that some star-forming spaxels (i.e. those not detected in CO by our ALMA observations) have gas fractions as low as retired spaxels, retired spaxels do not achieve gas fractions as high as those in star-forming spaxels.  

Our results are in agreement with the complementary studies of green valley galaxies by Lin et al. (2017) and Brownson et al. (2020) which find low central gas fractions and offsets from the rMGMS.  Indirect estimates of \sigh2, based on a calibration of the dust attenuation in a given spaxel, for larger samples of retired galaxies also find low gas fractions in the inner regions (Sanchez et al. 2018).  In addition to reduced gas fractions, there is also a reduction in the star formation efficiency (SFE) in retired/green valley/quenched galaxies (Sanchez et al., 2018; Brownson et al. 2020; Piotrowska et al. 2020; Colombo et al. 2020).  Future papers in the \aq\ series (Lin et al. in prep; Pan et al. in prep) will investigate the importance of both SFE and gas fraction in the green valley galaxies in the sample.

\section{Summary}

Using data obtained as part of the \aq\ survey, we combine optical IFU maps from MaNGA and molecular gas maps from ALMA to enable a study of the rMGMS for retired (rather than fully quenched) spaxels, which are characterized by photoionization by old stellar populations.  Our principal conclusions are:

\begin{itemize}
  
\item For the ensemble of spaxels in the 46 galaxies in the full \aq\ sample we demonstrate that, like star-forming spaxels, retired spaxels form a rMGMS.  However, the retired rMGMS is offset to lower \sigh2\ at fixed \sigstar\ and gas fractions are a factor of $\sim$5 lower for retired spaxels (Fig. \ref{MGMS_ensemble}).

\smallskip
  
\item  We select eight galaxies that 1) have axial ratio $b/a>$0.35, 2) contain at least 20 star-forming spaxels and 3) contain at least 20 retired spaxels.  The retired spaxels in 6/8 of these galaxies are centrally concentrated.  The remaining two galaxies have composite central regions with retired spaxels in a ring around them (Fig. \ref{map_fig}). Since, in all eight galaxies, the retired spaxels are at smaller radii than the star-forming spaxels, the cessation of star formation appears to occur from the inside-out.
  
\smallskip

\item  Both the star-forming and retired rMGMS show significant diversity in shape and normalization.  The retired rMGMS for a given galaxy forms a distinct sequence that is not simply a continuation of the galaxy's star-forming rMGMS to higher \sigstar\ (Fig. \ref{mgms_pifu}).

\smallskip

\item  Retired spaxels are more gas-poor, by up to an order of magnitude, than the star-forming spaxels in the same galaxy (Fig. \ref{fgas}).
  
\end{itemize}

Our results indicate that the cessation of star formation in the galaxies studied here is linked to a reduction (but not absence) of the molecular gas fraction on kpc-scales that is caused by a mechanism that works from the inside-out.   

\section*{Acknowledgements}

The authors acknowledge an NSERC Discovery Grant (SLE), AS Award CDA-107-M03 and MOST 108-2628-M-001 -001 -MY3 (LL), ERC Advanced Grant 695671 'Quench' and support from the STFC (AFLB).  This paper makes use of the following ALMA data:ADS/JAO.ALMA\#2015.1.01225.S, ADS/JAO.ALMA\#2017.1.01093.S, ADS/JAO.ALMA\#2018.1.00558.S, ADS/JAO.ALMA\#2018.1.00541.S.   ALMA is a partnership of ESO (representing its member states), NSF (USA) and NINS (Japan), together with NRC (Canada), MOST and ASIAA (Taiwan), and KASI (Republic of Korea), in cooperation with the Republic of Chile. The Joint ALMA Observatory is operated by ESO, AUI/NRAO and NAOJ.  The National Radio Astronomy Observatory is a facility of the National Science Foundation operated under cooperative agreement by Associated Universities, Inc.  Funding for the SDSS IV has been provided by the Alfred P. Sloan Foundation, the U.S. Department of Energy Office of Science, and the Participating Institutions. SDSS-IV acknowledges support and resources from the Center for High Performance Computing at the University of Utah. 

\section*{Data Availability}

The MaNGA data cubes used in this work are publicly available at https://www.sdss.org/dr15/.  The ALMA data used in this work are publicly available after the standard one year proprietary period via the ALMA archive: http://almascience.nrao.edu/aq/.

\end{document}